\documentclass[12pt,letter]{article}
\usepackage[isolatin]{inputenc}
\usepackage{listings}
\usepackage{xspace}
\usepackage{here}
\usepackage{graphics, graphicx}
\usepackage{amsmath}
\usepackage{citesort}

\usepackage{listings}

\def\eg{\emph{e.g.}\;}

\newcommand{\keyword}[1]{\text{\lstinline{#1}}\,}

\newcommand{\netTypek}{\keyword{Net}}

\newcommand{\lock}{\keyword{loc}}
\newcommand{\netk}{\keyword{net}}
\newcommand{\installk}{\keyword{install}}

\newcommand{\ink}{\keyword{in}}
\newcommand{\offk}{\keyword{off}}

\newcommand{\letk}{\keyword{let}}

\newcommand{\inactionn}{\offk}

\newcommand{\sensor}[5]{[{#1}::{#2}]^{{#3}, {#4}}_{{#5}}}
\newcommand{\sensord}{\sensor {\vec P} O p r e}
\newcommand{\tagsensor}[6]{[{#1}::{#2}]^{{#3}, {#4}}_{{#5}}\{{#6}\}}
\newcommand{\tagsensord}{\tagsensor {\vec P} O p r e S}

\newcommand{\moduled}{\{\method{l_i}{\vec x_i}{P_i}\}_{i\in I}}
\newcommand{\method}[3]{{#1} = ({#2})\,{#3}}

\newcommand{\invk}[3]{{#1}.{#2}({#3})}
\newcommand{\invkd}{\invk {v}{l}{\vec v}}

\newcommand{\install}[2]{{#1}.\installk {#2}}
\newcommand{\installd}{\install v v}

\newcommand{\Let}[3]{\letk {#1} = {#2}\, \ink {#3}}
\newcommand{\Letd}{\Let x P P}

\newcommand{\parn}{\,\vert\,}

\newcommand{\ldsq}{[\![} 
\newcommand{\rdsq}{]\!]} 
\def\context#1{\mathcal{C}\ldsq#1\rdsq}
\def\emptycontext{[\;]}

\def\obj#1{\{#1\}}
\def\objd{\obj {l_i \colon \vec T_i \rightarrow T_i}_{i \in I}}
\def\senObj#1{[#1]}
\def\senObjd{\senObj {l_i \colon \vec T_i \rightarrow T_i}_{i \in I}}
\def\anyObj#1{<#1>}
\def\anyObjd{\anyObj {l_i \colon \vec T_i \rightarrow T_i}_{i \in I}}
\newcommand{\dom}{\operatorname{dom}}
\newcommand{\disj}{\operatorname{,}}

\newcommand{\dist}{\operatorname{d}}

\newcommand{\fv}{\operatorname{fv}}

\newcommand{\energyIn}{\mathsf{e_{in}}}
\newcommand{\energyOut}{\mathsf{e_{out}}}
\newcommand{\methJoin}{+}
\newcommand{\subs}[2]{[{#1}/{#2}]}

\newcommand{\congr}{\equiv}
\newcommand{\reduces}{\rightarrow}

\newcommand{\disprule}[2]{
    #2 \tag{#1}
}

\newcommand{\mkRrule}[1]{{\footnotesize \textsc{R-#1}}}
\newcommand{\mkTrule}[1]{{\footnotesize \textsc{T-#1}}}
\newcommand{\mkSrule}[1]{{\footnotesize \textsc{S-#1}}}
\newcommand{\Rinstall}{\mkRrule{install}}
\newcommand{\RinstallTop}{\mkRrule{install-top}}
\newcommand{\Rlet}{\mkRrule{let}}
\newcommand{\Rswitch}{\mkRrule{switch}}
\newcommand{\Rmethod}{\mkRrule{method}}
\newcommand{\RmethodTop}{\mkRrule{method-top}}
\newcommand{\RnoMethodTop}{\mkRrule{no-method-top}}
\newcommand{\RnetCall}{\mkRrule{broadcast}}
\newcommand{\Rrelease}{\mkRrule{release}}

\newcommand{\Rnetwork}{\mkRrule{network}}

\newcommand{\Rcongr}{\mkRrule{congr}}

\newcommand{\SmonoidSensor}{\mkSrule{monoid-Sensor}}
\newcommand{\Sbroadcast}{\mkSrule{broadcast}}
\newcommand{\Sbattery}{\mkSrule{bat-exhausted}}

\newcommand{\type}{\vdash}
\newcommand{\TSinaction}{\mkTrule{term}}
\newcommand{\Tparallel}{\mkTrule{par}}
\newcommand{\TseqP}{\mkTrule{seqP}}
\newcommand{\TSsensor}{\mkTrule{sensor}}
\newcommand{\TSbSensor}{\mkTrule{bSensor}}

\newcommand{\Tlet}{\mkTrule{let}}
\newcommand{\Tcall}{\mkTrule{call}}
\newcommand{\Tinst}{\mkTrule{inst}}
\newcommand{\Tbcast}{\mkTrule{bcast}}
\newcommand{\TVvar}{\mkTrule{var}}

\newcommand{\TVseq}{\mkTrule{seq}}
\newcommand{\Tobj}{\mkTrule{obj}}
\newcommand{\Tbin}{\mkTrule{built-in}}
\newcommand{\Tnet}{\mkTrule{net}}
\newcommand{\Tloc}{\mkTrule{loc}}

\newcommand{\pad}{\;\;}
\newcommand{\Space}[1]{\pad{#1}\pad}
\newcommand{\grmeq}{\Space{::=}}

\newcommand{\grmor}{\;\mid\;}

\newenvironment{myfigure}{
  \begin{figure}[t]\centering\hrulefill\par\vspace{-4ex}}{
    \hrulefill\end{figure}}

\newcommand{\myparagraph}[1]{\vspace*{0.2cm}\noindent\textbf{#1}}

\newtheorem{theorem}{Theorem}

\newcommand{\rulespace}{0.0cm}

\begin{document}

\title{A Formal Model for Programming \\ Wireless Sensor Networks}
\author{Lu\'{\i}s Lopes$^1$, Francisco Martins$^2$, Miguel S. Silva$^1$, and Jo\~ao Barros$^1$\\
\small $^1$Departamento de Ci\^encia de Computadores, FCUP,  Portugal. \\
\small $^2$Departamento de Inform\'atica, FCUL, Portugal.
}

\date{}
\maketitle

\begin{abstract}
  In this paper we present new developments in the expressiveness and
  in the theory of a Calculus for Sensor Networks (CSN). We combine a
  network layer of sensor devices with a local object model to
  describe sensor devices with state. The resulting calculus is quite
  small and yet very expressive.  We also present a type system and a
  type invariance result for the calculus. These results provide the
  fundamental framework for the development of programming languages
  and run-time environments.
\end{abstract}

\textbf{keywords}: Sensor Networks, Ad-Hoc Networks, Ubiquitous
Computing, Process-Calculi, Programming Languages.

\section{Introduction}
\label{sec:introduction}

Developing an adequate programming model for sensor applications ---
involving highly dynamic networks with hundreds of power-constrained
and computationally restricted nodes~\cite{survey:akyildiz:etal:02}
--- stands out as a formidable goal.

A well developed programming model for sensor networks,
i.e.\@ a formalism or a calculus that captures their fundamental
computation and communication properties, and can serve as basis for
the development of higher level programming languages, is
likely to become a key enabler towards the aforementioned goal.

In the currently available implementations~\cite{culler:04}, the
sensor nodes are controlled by module-based operating systems such as
TinyOS~\cite{tinyos} and are programmed using somewhat ad-hoc
languages \eg nesC~\cite{nesc:gay:levis:etal} or
TinyScript/Mat\'{e}~\cite{mate:levis:culler:02}.  Recent middleware
developments such as Deluge~\cite{deluge:hui:culler:04} and
Agilla~\cite{agilla:fok:roman:lu:05} provide higher level programming
abstractions on top of TinyOS, including massive code deployment where
needed.

Our main contribution is a programming model with stronger formal
support and analytical capabilities than the above mentioned
solutions.  Beyond providing a rigorous representation (or a calculus)
of the sensor network at the programming level --- which allows for
formal verification of the correctness of programs and thorough
quantification of resource usage by sensors when running programs and
protocols --- our model provides a global vision of a sensor network
application, i.e. a specific distributed application, making it less
intuitive and error prone for programmers. Moreover, we do not require
the programs to be installed on each sensor individually, which would
be unrealistic for large sensor networks, allowing instead for dynamic
re-programming of the network.

Given the distributed and concurrent nature of sensor network
operations, we build our sensor network model based on process
calculi~\cite{pi:milner:parrow:walker:92,async-pi:honda:tokoro:91}
and also on an object
calculus~\cite{imperative-object-calculus:abadi:cardelli:95} to
introduce state into the sensors.
The associated theory is very rich and expressive.

Previous work on process calculi for wireless networks is scarce and
does not address the peculiarities of \emph{ad-hoc} sensor networks.
In~\cite{broadcast:prasad:91}, Prasad established the first process
calculus approach to modeling broadcast based systems.
Later work by Ostrovsk\'y, Prasad, and
Taha~\cite{broadcast-high-order:ostrovsky:prasad:taha:02} established
the basis for a higher-order calculus for broadcasting systems.
More recently, Mezzetti and
Sangiorgi~\cite{wireless:mezzetti:sangiorgi:06} discuss the use of
process calculi to model wireless systems.
The focus of this line of work lies in the protocol layer of the
networks.

In \cite{Silvaetal:07a} we presented a preliminary version of a
Calculus for Sensor Networks (CSN), devised as a two-layer calculus,
offering abstractions for data acquisition, communication, and
processing above the link layer of the protocol stack (i.e.~without
transmission errors and without packet losses). In this paper, we
offer several important improvements and complementary features, most
notably: (a) inclusion of state in CSN by combining it with a
fundamental object
calculus~\cite{imperative-object-calculus:abadi:cardelli:95}, leading
to a quite simple core calculus, and; (b) a type system for the
calculus and a \emph{subject reduction} result (types are invariant
under reduction).


\section{The Calculus}
\label{sec:calculus}

This section addresses the syntax and the semantics of the Calculus
for Sensor Networks. 
For simplicity, in the remainder of the paper we refer to a
sensor node or a sensor device in a network as a \emph{sensor}.
The syntax is provided by the grammar in
Figure~\ref{fig:syntax-sensor}, and the operational semantics is given
by the congruence and reduction relations depicted in
Figures~\ref{fig:structural-congruence} and~\ref{fig:reduction-semantics}.

\myparagraph{Syntax.}
\label{sec:syntax}
Let $\vec \alpha$ denote a possibly empty sequence $\alpha_1 \dots
\alpha_n$ of elements of the syntactic category $\alpha$.
Assume a countable set of \emph{labels}, ranged over by letter~$l$,
used to name methods within objects, and a countable set of
\emph{variables}, disjoint from the set of labels and ranged over by
letter $x$. 
Variables stand for communicated values (\emph{e.g.\@} basic values
and objects) in a given program context.

\begin{myfigure}
\begin{align*}
  & S \grmeq & & \text{\emph{Sensors}}                  & &  O \grmeq & & \text{\emph{Objects}} \\      
  & \qquad \; \inactionn & & \text{offline}           & &  \qquad \; \moduled & & \text{object API} \\
  & \quad \grmor \sensord & & \text{sensor}              & &  & & \\  
  & \quad \grmor S \parn S & & \text{composition}        & &  v \grmeq & & \text{\emph{Values}}\\ 
  & & &                                                 & &  \qquad \; b & & \text{built-in value}\\
  & P \grmeq & & \text{\emph{Programs}}                 & &  \quad \grmor x & & \text{variable}\\ 
  & \qquad \; v & & \text{value}                        & &  \quad \grmor \netk & & \text{broadcast}\\
  & \quad \grmor \invkd & & \text{method call}          & &  \quad \grmor \lock & & \text{sensor object}\\
  & \quad \grmor \installd & & \text{API update}        & &  \quad \grmor O & & \text{object} \\
  & \quad \grmor \Letd & & \text{local variable}        & &  & &
\end{align*}
\\[-0.4cm]
\caption{The syntax of CSN.}
\label{fig:syntax-sensor}
\end{myfigure}


%
%

A network is a flat, unstructured collection of
sensors $S$ combined using the parallel composition operator.
The sensors are assumed to be immersed in a (scalar or vector) field
and to be able to measure its intensity at their positions in space.
The field describes the distribution of some physical quantity we want
to model (\emph{e.g.\@} temperature, pressure, humidity) in space.
The position of a sensor is given in some coordinate system.
%

A sensor $\sensord$ represents an abstraction of a physical sensing
device running a sequence of instructions $\vec P$ and with a memory
$O$ (code plus state).
%
The object $O$ is a collection of methods, the API, that the sensor
makes available for internal and for external usage.
Each method, $l = (\vec x) P$, is identified by label $l$ and defined
by an abstraction $(\vec x) P$: a program $P$ with parameters $\vec
x$.
Method names are pairwise distinct within an object.
%
%
Intuitively, the collection of methods of a sensor may be interpreted
as the function calls of some tiny operating system installed in the
sensor at boot time or functionalities dynamically uploaded to the
sensor.
The position ($p$) of the sensors may vary with time if they are
mobile in some way.
The transmission range ($r$) on the other hand, usually remains
constant over time.
In this model we abstract away from battery ($e$) management. In the
operational semantics we simply check whether we have enough power for
certain operations (\eg broadcasting).


%

Programs are ranged over by $P$.
A method call, $\invkd$, calls the method~$l$ (with arguments
$\vec v$) in some value $v$. The value $v$ may be an anonymous object,
the sensor memory object, if the target is $\lock$, or the 
\emph{broadcast address}, if the target is $\netk$.
In the last case, the call is broadcasted to the network neighborhood
of the sensor.
Installing or replacing methods in an object can be done with 
the construct $\install v{v'}$, which adds the methods in $v'$ to 
the object in $v$, eventually replacing existing implementations 
with new ones. In particular, this construct allows the state of 
objects to be modelled.
The $\letk$ construct allows programs to create local variables to
hold intermediate values in computations. In particular, it allows the
construction of arbitrarily complex data structures when combined with
the appropriate methods in the sensor object.

We do not have a primitive sequential composition construct for
programs. Such a construct can be easily obtained as syntactic sugar
as: $\letk\ x = P\ \ink\ Q \equiv P;Q$ where $x \not\in \fv(P)$. The
semantics of the calculus forces the evaluation of $P$ first and then
$Q$ exactly, since $x$ does not occur free in $Q$.
Thus, although we do not have a primitive sequential composition
construct in the calculus, in the remainder of this paper we use this
construct to impose a more imperative style of programming.

Values are the data exchanged between sensors and comprise basic
values that can intuitively be seen as the primitive data types
supported by the sensor's hardware, and objects that are constructed
dynamically.
Notice that this is not a higher-order calculus: we can only transfer
the \emph{code}, retransmit, or install objects in remote sensor.

\myparagraph{A Simple Example.}
\label{sec:simple-example}
We start with a very simple \lstinline{ping} program.
We denote as \lstinline{MSensor} and \lstinline{MSink} the
objects installed in any of the anonymous sensors in the network and
in the sink, respectively.
Each sensor has a \lstinline{ping} method that when called broadcasts
a \lstinline{forward} call to the network with its MAC
address~\lstinline{m}, and broadcasts another \lstinline{ping} call to
propagate the call in the network.
%
%
The sink has a distinct implementation of this method.
Any incomming call logs the MAC address given as argument.
So, the overall result of the call \lstinline{net.ping()} in the sink
is that all reachable sensors in the network will, in principle,
receive this call and will flood the network with their MAC addresses.
These values eventually reach the sink and get logged.

\begin{lstlisting}
MSensor(m) = { ping    = ()  net.forward(m); net.ping()
               forward = (x) net.forward(x) }
MSink      = { forward = (x) log_mac(x) }
[net.ping(), MSink] | [{}, MSensor(mx)] | ... | [{}, MSensor(my)] 
\end{lstlisting}

\myparagraph{Semantics.}
\label{sec:semantics}
%
The calculus has two variable bindings: the $\letk$ construct and method 
definitions.
The displayed occurrence of variable $x$ is a \emph{binding} with
\emph{scope}~$P$ both in $\letk\ x=P'\ \ink\ P$ and in $\method l
{\dots, x, \dots} P$.
An occurrence of a variable is \emph{free} if it is not in the scope of a 
binding.
Otherwise, the occurrence of the variable is \emph{bound}. 
The set of free variables of a sensor $S$ is referred as $\fv(S)$.

\begin{myfigure}
  \begin{gather*}
     \tag{\SmonoidSensor}
    S_1 \parn S_2 \congr S_2 \parn S_1,
    \qquad
    S \parn \inactionn \congr S
    \qquad
    S_1 \parn (S_2 \parn S_3)  \congr (S_1 \parn S_2) \parn S_3
    \\
    \tag{\Sbroadcast, \Sbattery}
    \sensord \congr \tagsensor{\vec P}{O}{p}{r}{e}{\inactionn}
    \qquad \qquad
    \frac{
      e < \min(\energyIn, \energyOut)
    }{
      \sensor {\vec P}{O}{p}{r}{e} \congr \inactionn
    }
  \end{gather*}
  \\[-0.4cm]
\caption{Structural congruence for processes and sensors.}
\label{fig:structural-congruence}
\end{myfigure}


Following Milner~\cite{ccs:milner:80} we present the reduction relation 
with the help of a structural congruence relation.
The structural congruence relation $\congr$, depicted in
Figure~\ref{fig:structural-congruence}, allows for the manipulation of
the syntactic structure of terms, making it possible for sub-terms to
reduce.
The relation is defined as the smallest congruence relation on sensors
closed under the rules given in Figure~\ref{fig:structural-congruence}.

The parallel composition of sensors is commutative and associative
with $\inactionn$ as the neutral element (\emph{vide} Rule
\SmonoidSensor).
%
%
%
When a sensor is broadcasting a message it uses a conceptual
\emph{membrane} to engulf the sensors as they become engaged in
communication. Rule \Sbroadcast{} allows for a sensor to start the
broadcasting operation.
An offline sensor is one with insufficient battery capacity for
performing an internal or an external reduction step (\emph{vide} Rule
\Sbattery).

The reduction relation on networks, notation $S \reduces S'$,
describes how a sensor $S$ can evolve (reduce) to sensor $S'$.
Reduction in a sensor occurs at the head of a sequence $\vec P$. 
In other words, in a sequence $P,\vec P$, the program~$P$ is running
and those in $\vec P$ are waiting in a queue.
However, we also allow for reduction within the $\letk$ construct.
In other words, the $P$ in the example above can be of the form $\letk\
x=P'\ \ink\ P''$ and we allow the reduction \emph{in situ} of $P'$.
Naturally, we may have multiple levels of $\letk$ constructs involved.
For this reason we present our reduction relation using reduction
contexts, or places were reduction may occur.
These contexts, denoted $\context{}$, are defined as follows:
\[
  \context{} \grmeq \emptycontext \grmor \letk\ x=\emptycontext\ \ink\ P
\]
Thus, $\context{P}$ denotes the process $P$ inserted in the
$\emptycontext$ hole of any of the above contexts.

The reduction relation is inductively defined by the rules in
Figure~\ref{fig:reduction-semantics}.
The reduction for sensors is parametric on two constants 
$\energyIn$ and $\energyOut$ that represent the amount of
energy consumed when performing internal computation steps
($\energyIn$) and when broadcasting messages ($\energyOut$).

\begin{myfigure}
  \begin{gather*}
    \tag{\RmethodTop}
    \frac{
       O(l) = (\vec x) P
       \qquad
       e \geq \energyIn
    }{
       \sensor {\context {\lock.l[\vec v]}, \vec P}{O}{p}{r}{e}
       \reduces
       \sensor{\context {P \subs {\vec v} {\vec x}}, \vec P}{O}{p}{r}{e}
    }
    \\[\rulespace]
    \tag{\RnoMethodTop}
    \frac{
      l \not \in \dom(O)
    }
    {
      \sensor {\context {\lock.l[\vec v]}, \vec P} O p r e 
      \reduces
      \sensor {\context {\lock.l[\vec v]}, \vec P} O p r e
    }
    \\[\rulespace]
    \tag{\Rmethod}
    \frac{
      O' (l) = (\vec x) P
      \qquad
      e \geq \energyIn
    }{
      \sensor {\context{O'.l[\vec v]}, \vec P} O p r e 
      \reduces
      \sensor {\context{P \subs {\vec v} {\vec x}}, \vec P} O p r e 
    }
    \\[\rulespace]
    \disprule{\RnetCall}
    {
      \frac{
        \dist (p, p') < r 
        \qquad
        e \geq \energyOut
      }
      {
        \begin{array}{l}
        \!\!\!
        \tagsensor{\context{\invk \netk {l_i} {\vec v}}, \vec P} O p r e S \parn
        \sensor{\vec P'}{O'}{p'}{r'}{e'}  
        \reduces\\
        \hspace{1.7cm}
        \tagsensor {\context{\invk \netk {l_i}{\vec v}}, \vec P}  {O} {p} {r} {e} 
        {S \parn \sensor{\vec P', \invk {\lock} {l_i} {\vec v}}{O'}{p'}{r'}{e'}}
        \!\!\!
        \end{array}
      }
    }
    \\[\rulespace]
    \disprule{\Rrelease}
     {
        \tagsensor {\context{\invk \netk {l_i} {\vec v}}, \vec{P}} O p r e S
        \reduces
        \sensor {\context{\obj{}},\vec P} O p r e \parn S
    }
    \\[\rulespace]
    \tag{\RinstallTop}
    \frac{
      e \geq \energyIn
    }
    {
      \sensor {\context{\install \lock {O'}}, \vec P} O p r e
      \reduces
      \sensor {\context{O \methJoin O'}, \vec P} {O \methJoin O'} p r e
    }
    \\[\rulespace]
    \tag{\Rinstall}
    \frac{
      e \geq \energyIn
    }
    {
      \sensor {\context{\install {O'} {O''}}, \vec P} O p r e
      \reduces
      \sensor {\context{O' \methJoin O''}, \vec P} O p r e
    }
    \\[\rulespace]
    \tag{\Rlet}
    \frac{
      e \geq \energyIn
    }
    {
      \sensor {\context{\Let x v P}, \vec P} O p r e 
      \reduces
      \sensor {\context{P \subs v x}, \vec P} O p r e
    }
    \\[\rulespace]
    \tag{\Rswitch, \Rnetwork}
    \sensor {P, \vec P} O p r e
    \reduces
    \sensor {\vec P, P} O p r e
    \qquad
    \frac{
      S \reduces S'
    }
    {
      S \parn S''
      \reduces
      S' \parn S''
    }
    \\[\rulespace]
    \tag{\Rcongr}
    \frac{
      S_1 \congr S_2 
      \qquad
      S_2 \reduces S_3
      \qquad
      S_3 \congr S_4
    }{
      S_1 \reduces S_4
    }
\end{gather*}
\\[-0.4cm]
\caption{Reduction semantics for sensors.}
\label{fig:reduction-semantics}
\end{myfigure}


A program $P$ in a sensor $\sensord$ may: (a) call 
methods in the top level object $O$ (Rules \RmethodTop\ and 
\RnoMethodTop), in anonymous objects (Rule \Rmethod), and in 
the network neighborhood (Rules \RnetCall{} and \Rrelease); 
(b) install new methods in the top level object $O$
(Rule \RinstallTop), and in anonymous objects (Rule \Rinstall);
(c) compute intermediate values and assign them to new variables 
(Rule \Rlet), and (d) stop and allow another program to run 
(Rule \Rswitch).

A call to a local method $l$ with arguments $\vec v$ in an
object $O$, be it the top level object or an anonymous one, such that
$O(l)=(\vec x)P$, results in the program $P$ where the variables in
$\vec x$ are replaced with the values $\vec v$.
Traditionally, typed programming languages use a type system to ensure
that there are no calls to undefined methods, ruling out all
other programs at compile time.
Our approach allows an extra degree of flexibility.
When the method $l$ is not present in object $O$ the reduction depends
on whether $O$ is the top level object or not.
In the first case we have decided to keep the call active
(see Rule \RnoMethodTop).
In the latter, calling an undefined method in an anonymous object
causes the program to get \emph{stuck}.
%
%
%
%
We envision that if we call a method in the network after some code
has been deployed (see Section~\ref{sec:examples}), some sensors may 
receive the method call before the code is actually 
deployed.
With the semantics we propose, the call actively waits for the code to
be installed.

Sensors communicate with the network by broadcasting messages. 
A message consists of a remote method call on unspecified
sensors in the neighborhood of the emitting sensor.
In other words, the messages are not targeted to a particular sensor
(there is no peer-to-peer communication).
The neighborhood of a sensor is defined by its communication radius,
but there is no guarantee that a message broadcasted by a given sensor
arrives at all surrounding sensors.
%
%
Also, during a broadcast operation the message must only reach
each neighborhood sensor once.
Notice that we are not saying that the same message can not reach the
same sensor multiple times.
In fact it might, but as the result of the echoing of the message in
subsequent broadcast operations.
We model the broadcasting of messages in two stages.
Rule \RnetCall{} calls method $l$ in the remote sensor, provided
that the distance between the emitting and the receiving sensors
is less that the transmission radius ($\dist(p, p') < r$).
Each sensor that receives the call is put in the membrane associated
with the emitting sensor, thus preventing multiple deliveries of the
same message while broadcasting.
Observe that the rule does not enforce the interaction with all
sensors in the neighborhood of the emmiting sensor.
Rule \Rrelease{} finishes the broadcast by consuming the operation
($\invk \netk {l} {\vec v}$), and dissolves the membrane.
A broadcast operation starts with the application of Rule~\Sbroadcast,
proceeds with multiple (eventually none) applications of
Rule~\RnetCall{} (one for each target sensor), and terminates with the
application of Rule~\Rrelease.

Installing a set of methods $O'$ in an existing object $O$, be it at
top level or not, amounts to adding to $O$ the methods in $O'$ (absent
in $O$), and to replace (in $O$) the methods common to both $O$ and
$O'$.
Rigorously, the operation of installing the methods $O'$ on top of $O$,
denoted $O \methJoin O'$, may be defined as $O \methJoin O' = (O
\setminus O') \cup O'$.
The $\methJoin$ operator is reminiscent of Abadi and Cardelli's
operator for updating methods in their imperative object
calculus~\cite{imperative-object-calculus:abadi:cardelli:95}.

A running program $P$ in a sequence $P,\vec P$, may stop its 
execution at any time and allow the execution of the next program 
in the sequence $\vec P$.
Program $P$ is placed at the end of the sequence, actually
implementing a very simple \emph{round-robin} interleaved execution
model.
This feature is essential to ensure that sensors eventually process
incomming network communication.
Intuitively, this rule may be seen as a very simple scheduling
mechanism provided by an underlying operating system in each sensor.
 


\section{Further Examples}
\label{sec:examples}

In this section we present some further examples, programmed in CSN, of 
operations performed on sensor networks.
%
%
%
%
Finally, we assume in these examples that the network layer 
supports \emph{scoped flooding}.
Software based \emph{scoped flooding} can be easily implemented with
CSN although the code is a bit lengthy to present here.

\myparagraph{Polling.} In this example the sink instructs the nodes
to sample the field continuously and logs the results.
The sink just calls the method \lstinline{sample} once on the
network.
This method propagates the call through the network and calls
\lstinline{sample}, for each sensor.
The first call to \lstinline{sample} starts by propagating the call to
the network neighborhood, changes itself through an \installk{}
call and finally calls itself to start the sampling cycle.
The newly installed code of \lstinline{sample} reads values from the
field, forwards the results to the network, and calls itself again 
to implement a cycle.

\begin{lstlisting}
MSensor   = {
  sample  = ()  net.sample();
          install {sample = () 
             let f = loc.field() in net.forward(f);loc.sample()};
          loc.sample()
  forward = (x) net.forward(x)
}
MSink     = { forward = (x) log_field(x) }
[net.sample(), MSink] | [{}, MSensor] | ... | [{}, MSensor] 
\end{lstlisting}

\myparagraph{Code Deployment.} The above example assumes we have some
means of deploying the code to the sensors.
In this example we address this problem and show how it can be
programmed in CSN.
The code we wish to deploy and execute is the same as the one in the
previous example.
Here, we deploy the code in the network by sending an object with methods 
\lstinline{sample} and \lstinline{forward} to all the sensors.
We do this by calling \lstinline{deploy} on the network and sending
the above mentioned object as a parameter.
Once deployed, the code is activated with a call to \lstinline{sample}
from the sink as above.

\begin{lstlisting}
MSensor   = { deploy  = (x) install x; net.deploy(x) }
MSink     = { forward = (x) log_field(x) }
[net.deploy[{
  sample  = ()  net.sample();
          install {sample = () 
            let f = loc.field() in net.forward(f); loc.sample()};
          loc.sample()
  forward = (x) net.forward(x)
 }]; net.sample(), MSink] | [{}, MSensor] | ... | [{}, MSensor]
\end{lstlisting}

\section{The Type System}
\label{sec:types}

In this section we present a simple type system for CSN and informally
discuss runtime errors.

The syntax for types is depicted in Figure~\ref{fig:syntax-types}.
Types $T$ are built from the built-in type $B$ and the broadcast type $\netTypek$
using the constructor for object types $\objd$ (and $\senObjd$).
Type $B$ is the type of built-in values (\eg battery, position, energy).
Type $\netTypek$ is assigned to value $\netk$ and distinguishes local
method invocations from broadcast communications (via remote method
invocation).
A type $\objd$ describes an object represented as a collection
of (distinctly named) methods.
Each method $l_i$ has type $\vec T_i \rightarrow T_i$, where
$\vec T_i$ is the type of the parameters of the method
and~$T_i$ is its return type.
For instance, type $\obj {\text{\lstinline{ping}}\colon \epsilon
  \rightarrow \obj{}, \text{\lstinline{forward}}\colon B \rightarrow
  \obj{}}$ is the type of the object \lstinline{MSensor} presented in
Section~\ref{sec:calculus}.
It represents an object with two methods named \lstinline{ping} and
\lstinline{forward}. 
Method \lstinline{ping} has no parameters and returns an empty
object (the result from the final \lstinline{net.ping()} operation).
Method \lstinline{forward} accepts a built-in value and  
returns $\obj{}$, like \lstinline{ping}.
Type $\senObjd$ denotes the type of the sensor's object.
It plays an important role when typing code installation.
When referring to both types of objects we use notation
$\anyObjd$.

\begin{myfigure}
\begin{align*}
  & T \grmeq & & \text{\emph{Types}}
  \\
  & \qquad \; B & & \text{built-ins} & &
                        \quad \grmor  \netTypek & & \text{broadcast}
  \\
  & \quad \grmor  \objd & & \text{objects} & &
                        \quad \grmor  \senObjd & & \text{sensor object}
\end{align*}
\\[-0.4cm]
\caption{The syntax of types.}
\label{fig:syntax-types}
\end{myfigure}


\begin{myfigure}
  \begin{gather*}
    \Gamma \type b \colon B
    \qquad
    \Gamma \disj x \colon T \type x \colon T
    \qquad
    \frac{
      \forall i. \Gamma \type x_i \colon T_i
    }
    {
      \Gamma \type \vec x \colon \vec T
    }
    \tag{\Tbin, 
      \TVvar, \TVseq}
    \\[\rulespace]
    \Gamma \type \netk \colon \netTypek
    \qquad
    \Gamma \disj \lock \colon \senObjd \type \lock \colon \senObjd
    \tag{\Tnet, \Tloc}
    \\[\rulespace]
    \frac{
      \forall i \in I.\Gamma \disj \vec x_i \colon \vec T_i \type P_i \colon T_i
    }
    {
      \Gamma \type \moduled \colon \objd
   }
   \tag{\Tobj}
 \end{gather*}
 \\[-0.4cm]
\caption{Typing rules 
  for values.}
\label{fig:type-system-values}
\end{myfigure}


\begin{myfigure}
  \begin{gather*}
    \tag{\Tcall}
    \frac{
      \Gamma \type v \colon \anyObjd
      \qquad
      \Gamma \type \vec v \colon \vec T_j
      \qquad
      j \in I
    }
    {
      \Gamma \type \invk v {l_j} {\vec v} \colon T_j
    }
    \\[\rulespace]
    \tag{\Tbcast}
    \frac{
      \Gamma \type v \colon \netTypek
      \qquad
      \Gamma \type \invk \lock l {\vec v} \colon \_
    }
    {
      \Gamma \type \invkd \colon \obj {}
    }
    \\[\rulespace]
    \tag{\Tinst}
    \frac{
      \Gamma \type v_1 \colon T_1
      \qquad
      \Gamma \type v_2 \colon T_2
      \qquad
      T_1 \oplus T_2 \text{ defined}
    }
    {
      \Gamma \type \install {v_1} {v_2} \colon T_1 \oplus T_2
    }
    \\[\rulespace]
    \tag{\Tlet}
    \frac{
      \Gamma \type P_1 \colon T_1
      \qquad
      \Gamma
      \disj x \colon T_1 \type P_2 \colon T_2
    }
    {
      \Gamma \type \Let x {P_1} {P_2} \colon T_2
    }
  \end{gather*}
  \\[-0.4cm]
\caption{Typing rules for programs.}
\label{fig:type-system-programs}
\end{myfigure}


\begin{myfigure}
  \begin{gather*}
   \frac{
     \begin{array}{c}
     \Gamma
     \type O \colon \objd
     \qquad
     \Gamma
     \type \vec P \colon \_
     \qquad
     \Gamma \type p r e \colon \vec B
     \\
     \Gamma \type \lock \colon \senObj {l_j \colon \vec T_j \rightarrow T_j}_{j \in J}
     \qquad
     I \subseteq J
   \end{array}
   }
   {
     \Gamma \type \sensord
   }
   \tag{\TSsensor}
   \\[\rulespace]
   \Gamma \type \inactionn
   \qquad
   \frac{
     \Gamma \type \sensord
     \qquad
     \Gamma \type S
   }
   {
     \Gamma \type \tagsensord 
   }
   \tag{\TSinaction, \TSbSensor}
   \\[\rulespace]
   \frac{
     \Gamma \type S_1 
     \qquad
     \Gamma \type S_2
   }
   {
     \Gamma \type S_1 \parn S_2
   }
   \qquad
   \frac{
     \Gamma \type P \colon \_
     \qquad
     \Gamma 
     \type \vec P \colon \_
   }
   {
     \Gamma \type P, \vec P \colon \_
   }
   \tag{\Tparallel, \TseqP}
 \end{gather*}
 \\[-0.4cm]
\caption{Typing rules for sensor networks and for program sequences.}
\label{fig:type-system-sensors}
\end{myfigure}


The type system is defined in Figures~\ref{fig:type-system-values},
\ref{fig:type-system-programs},
and~\ref{fig:type-system-sensors}.
A typing $\Gamma$ is a partial function of finite domain from variables
to types.
We write $\dom(\Gamma)$ for the domain of $\Gamma$.
When $x \not \in \dom(\Gamma)$ we write $\Gamma, x \colon T$ for the 
typing $\Gamma'$ such that $\dom(\Gamma') = \dom(\Gamma) \cup \{x\}$,
$\Gamma'(x) = T$, and $\Gamma'(y) = \Gamma(y)$ for $y \neq x$.

Type judgements are of three forms: $\Gamma \type v \colon T$ means
that value $v$ has type~$T$, under the assumptions in typing $\Gamma$;
$\Gamma \type P \colon T$ asserts that program~$P$ has type $T$,
under the assumptions in $\Gamma$; and $\Gamma \type S$ means that 
sensor network~$S$ is well typed, assuming the typing $\Gamma$.

The rules for typing values (Figure~\ref{fig:type-system-values})
are straightforward. 
Rule $\Tloc$ assigns type $\senObjd$ to the (special) variable $\lock$.
It represents the interface of all sensors and is invariant during
type checking. 
This means that the interface for sensors is fixed by the
application programmer, before type checking takes place.

As for programs, method calls are separated in local calls (rule
\Tcall) and remote calls (rule \Tbcast).
In a local call, method $l_j$ must be part of the target object ($j
\in I$), the type of the arguments must agree with the type of the
parameters ($\vec v \colon \vec T_j$), and the type of the invocation
($T_j$) is the return type of the method.
A remote call ($v \colon \netTypek$) is type checked as a local call,
apart from its return type that is always the empty object ($\obj{}$),
meaning that the return value of a remote call is ignored.
Code installation (rule \Tinst) is allowed either in anonymous objects,
or in the sensor's object.
The definition of operation $T_1 \oplus T_2$ is similar to that of
operation $+$ for combining objects, but has only a meaning for
$\senObj{\_} \oplus \senObj{\_}$, $\senObj{\_} \oplus \obj{\_}$, and
$\obj{\_} \oplus \obj{\_}$.
Allowing $\obj{\_} \oplus \senObj{\_}$ may cause an anonymous object
to refer to undefined methods, since it inherits the complete interface
from the type of the top level object.
The result of an $\installk$ operation is the altered object.

Regarding the typing rules for sensors, we focus on rule \TSsensor,
as the remainder of the rules should be simple to follow.
When typing a sensor we make sure that the methods available in the
sensor's object conform with the global, network-wide, interface
defined by $\lock$.
Notice that a sensor may offer just a subset of the interface methods
($I \subseteq J$), since some of them may not yet be available
(installed) in the sensor.

The following result ensures that types are preserved during
reduction.
\begin{theorem}[Subject Reduction]
  If $\Gamma \type S$, $S \reduces S'$, then $\Gamma \type S'$.
\end{theorem}
The proof proceeds by induction on the derivation tree for the
reduction $S \reduces S'$ and is a straightforward case analysis; due
to space constraints we omit it, as well as the standard intermediate
results required for the proof.


\section{Conclusions and Future Work}
\label{sec:conclusions}

Based on our formal computational model for programming
sensor networks (presented in~\cite{Silvaetal:07a}), 
we developed a natural extension to account for sensor nodes
with different states. In addition, we introduced a static type 
system that enables safe programming of sensor networks. 
It is worth pointing out that typed sensor network applications can be filtered
at compile time, allowing for premature detection of  
certain types of programs that would produce run-time errors. 
As a first step towards proving the \emph{type safeness} of the model, we
provided a \emph{subject reduction} result.

It is our belief that these results set the basis for a range of programming 
language idioms for sensor networks which we aim to implement in the 
immediate future. Ultimately, we would like to program real-world sensor
network applications using such a language and its associated
run-time system.



\bibliography{refs}
\bibliographystyle{plain}

\end{document}